%% file: NA_Main.tex
\documentclass[prl,reprint,superscriptaddress]{revtex4-1}

\usepackage{amsfonts}           
\usepackage{amssymb}            
\usepackage{amsmath}            
\usepackage{latexsym}           
\usepackage{dcolumn}            
\usepackage{enumerate}
\usepackage{epstopdf}           
\usepackage{fancyhdr}           
\usepackage[pdftex]{graphicx}   
\usepackage{hyperref}           
\usepackage{setspace}           
\usepackage{xspace}             
\usepackage{subfigure}          
\usepackage{bm}                 
\usepackage[usenames]{color}      
\usepackage{multirow}           

\newcommand{\bZ}{\mathbb{Z}}
\newcommand{\br}{\mathbf{r}}

\newcommand{\zTwo}{\mathbb{Z}_2\times\mathbb{Z}_2}

\newcommand{\aFive}{A_5}

\definecolor{zachCommentRed}{rgb}{1.0,0.211,0.211}
\definecolor{zachSaveGreen}{rgb}{0.211,1.0,0.211}
\definecolor{zachLaterBlue}{rgb}{0.211,0.211,1.0}
\newcommand{\comment}[1]{\textcolor{zachCommentRed}{}}
\newcommand{\save}[1]{\textcolor{zachSaveGreen}{}}
\newcommand{\later}[1]{\textcolor{zachLaterBlue}{}}
\newcommand{\change}[1]{\textcolor{zachCommentRed}{}}

\begin{document}

\title{Classical Topological Order in Abelian and Non-Abelian Generalized Height Models}
\date{\today}                                                  
\author{\firstname{R. Zach} Lamberty}
\affiliation{LASSP, Cornell University, Ithaca, New York, 14853}      
\author{Stefanos Papanikolaou}
\affiliation{LASSP, Cornell University, Ithaca, New York, 14853}      
\affiliation{Dept. of Physics, Yale University, New Haven, Connecticut, 06520}
\author{\firstname{Christopher L.} Henley}
\affiliation{Cornell University, Ithaca, New York, 14853}      

\begin{abstract}
We present Monte Carlo simulations on a new class of lattice models in which the degrees of freedom are elements of an abelian or non-abelian finite symmetry group $\mathcal{G}$, placed on directed edges of a two-dimensional lattice. The plaquette group product is constrained to be the group identity. In contrast to discrete gauge models (but similar to past work on height models) only elements of symmetry-related subsets $\mathcal{S}\in\mathcal{G}$ are allowed on edges. These models have topological sectors labeled by group products along topologically non-trivial loops. Measurement of relative sector probabilities and the distribution of distance between defect pairs are done to characterize the types of order (topological or quasi-LRO) exhibited by these models.  We present particular models in which \emph{fully local} non-abelian constraints lead to \emph{global} topological liquid properties.
%
%
%
%
\end{abstract}
\maketitle

\input{NA_Intro.tex}

\input{NA_TheModel.tex}

\input{NA_SecProb.tex}

\input{NA_DefPairDist.tex}
\input{NA_Conclusions.tex}

\emph{\underline{Acknowledgements}}: This work was supported by the National Science Foundation through a Graduate Research Fellowship to R. Zach Lamberty and grant DMR-0552461.


\end{document}

%% file: NA_Intro.tex
\emph{\underline{Introduction}}: Ever since Wen first proposed that ``Topological Order" be considered as a means of classifying chiral spin states in superconductors \cite{XGWen89}, applications of this ``post-Landau" paradigm have caused great excitement in applications to {\it quantum} systems---the integer \cite{ThoulessKohmotoNightingale82} and fractional \cite{XGWen90} quantum Hall effect, spin liquids\cite{KalmeyerLoughlin,WenWilczekZee,ReadSachdev,XGWen91}, and topological quantum computation \cite{Kitaev03}.

Most other physical phenomena, such as critical phenomena, were understood classically prior to being realized in quantum mechanics. The question arises whether there is a useful notion of topological order in a {\it classical} setting. Let us define a generalized topological order in a classical ensemble by the existence of sectors of (energetically or entropically) degenerate states, disconnected in the thermodynamic limit, which cannot be distinguished by any local order parameter. One can realize this sort of order in classical models where the inaccessibility is built in by hand, or in a limit where inter-sector transitions are suppressed by an activation energy much larger than the temperature \cite{CastelnovoChamon07,Henley11}. The motivation for such models is that first, by sharing many features with the quantum models, they provide an arena to do calculations that would not be feasible in the quantum case; secondly, such a classical ensemble can furnish the Hilbert space for contructing a quantum model with the same topological order (e.g. the classical $\mathbb{Z}_2$ topological order in dimer coverings on triangular lattices \cite{Moessner01}).

As a particular instance of classical topological ordering, one of us has proposed \cite{Henley11} a new family of models which we call ``generalized height models" which are based on an abelian or non-abelian {\it discrete} group in the same way that lattice ``height'' models \cite{vanBeijeren77,BloteHilhorst82,ZhengSachdev94,Levitov90,KondevHenley95} are based on the integers. These models share many properties of their quantum analogues, including massively degenerate ground states, topological defect charges, and the possibility in non-abelian cases of combining two defect charges in more than one way (called ``fusion channels" in the quantum context). \save{CLH: nice here to cite a review article on TO but not vital.} In this paper we present simulation results from these models, focused primarily on two measurements which characterize whether a given model is topologically ordered: the relative probabilities of being each sector, and the distribution of separations between a pair of topological defects.

%% file: NA_TheModel.tex

\emph{\underline{Simulations of the Model}}: 
Generalized height models can be thought of as lying between (i) ``height models''~\cite{vanBeijeren77,BloteHilhorst82,ZhengSachdev94,Levitov90,KondevHenley95}, in which directed edges are labeled by the differences between integer-valued heights) on sites; and (ii) lattice gauge models~\cite{DoucoutIoffe05} based on a finite group, in which any group element (which we shall call a ``spin") may be assigned to each (directed) lattice edge.

Given a discrete group $\mathcal{G}$, we define a particular generalized height model as the (massively degenerate) ensemble of all configurations satisfying two local constraints. First, the plaquette constraint
\begin{equation}
    \prod_{i\in\textit{Plaq}}\sigma(\br_i,\br_{i+1}) = e,
    \tag{C1}
    \label{eq:cPlaq}
\end{equation}
where $\sigma(\br_i,\br_{i'})$ is the ``spin'' placed on the edge between vertices $i$ and ${i'}$, the product runs around the four edges of a plaquette, and $e$ is the group identity (this generalizes the constraint in a height model that height differences add to zero around a plaquette). Wherever constraint \eqref{eq:cPlaq} is violated, that plaquette is said to have a ``charge'' defect, with its charge being the plaquette product.

\begin{figure}[!h]
    \includegraphics[width = 1.75 in]{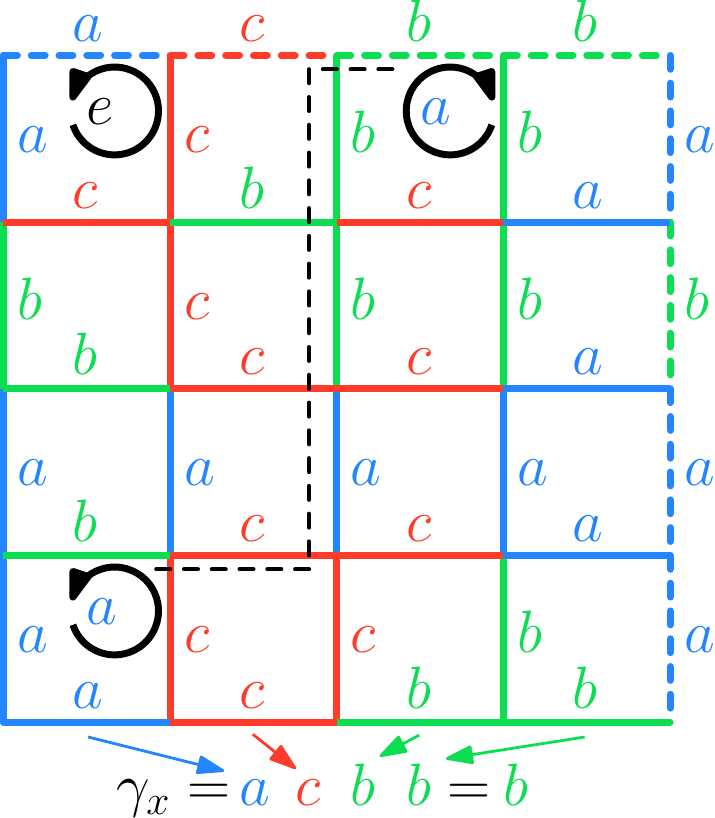}
    \caption{\label{fig:examplePlaquette} A sample configuration of our model with group $\mathcal{G}=\zTwo$ on a square lattice with periodic boundary conditions. The group elements are the identity $e$ and $\{a,b,c\}$ with multiplication table $a^2 = b^2 = c^2 = e, ab=ba=c, bc=cb=a, ca=ac=b$ (this $a,b,c$ notation is used throughout the paper); edges are occupied by any of $\{a,b,c\}$. On all but two plaquettes (\textit{e.g} upper-left corner) the product of group elements along the edges is $e$. The two plaquettes connected by a black dashed line are defect plaquettes with opposite ``charges," on which the plaquette product is not the identity. We have also shown $\gamma_x$, the plaquette product along one of the topologically non-trivial loops, which is the first component of the ``sector label" for this configuration.}
\end{figure}

Secondly, in contrast to discrete gauge models (but like height models), \emph{only} elements of a chosen symmetry-related strict subset
\begin{equation}
    \mathcal{S}\subsetneq\mathcal{G}
    \tag{C2}
    \label{eq:cSpin}
\end{equation}
are placed on the directed edges of the lattice. The set $\mathcal{S}$ consists of one or sometimes more symmetry classes of group elements. An example of a generalized height model can be seen in figure \ref{fig:examplePlaquette}.

In discrete gauge models, \textit{any} group elements is allowed on an edge and therefore the spin-spin correlation vanishes (\textit{e.g.} as in the toric code \cite{Kitaev06}). A generalized height model generally has exponentially decaying spin correlations. By changing the size of the allowed spin subset we can discretely interpolate between totally free gauge models and quasi-long range ordered states, as elaborated elsewhere \cite{LambertyPapnikolaouHenley}.

Constraint \eqref{eq:cPlaq} implies the product of ``spins'' along any topologically trivial loop in the lattice must be the identity, but along a topologically non-trivial loop it may be another group element. Our simulations take place on the torus, and we adopt the two independent non-trivial loop products as labels for the disjoint partitions of our ensemble, hereafter referred to as \textit{sectors}.  In the case of abelian $\mathcal{G}$, each sector has a uniquely defined label; for non-abelian groups sectors are defined up to conjugacy \cite{Henley11}.\save{Specifically, group products along neighboring paths, $\gamma_0$ and $\gamma_1$, are related to each other by conjugacy with the elements on the lattice edges $s$ connecting them, $\gamma_0 = s^{-1}\gamma_1 s$.)} Local updates cannot take us from one sector to another, but global updates can.

Our Monte Carlo simulations use a sequence of local and global update moves (satisfying detailed balance). The local move is a single-vertex update in which we multiply all outgoing edges for one vertex by some group value $g\in\mathcal{G}$; this preserves the plaquette and sector products, but must be rejected if any of the resulting ``spins'' on the outgoing edges are not in the allowed set $\mathcal{S}$.

Our chosen global move involves creating a defect/anti-defect pair, randomly walking one of the defects around the torus, and letting the pair recombine into a new defect (only possible in non-Abelian models) or annihilate.\save{Each individual jump of a defect from one plaquette to the next is referred to as a defect step, and the entire process is called a completed defect walk.}

If a defect walks once around some topologically non-trivial loop, every transverse loop is crossed once and has its loop product changed, so that the new configuration resides in a different sector. Such updates, that change sectors while satisfying detailed balance, allow us to measure the relative weights of the different sectors and thereby ascertain whether the ensemble is topologically ordered.

Constraint \eqref{eq:cSpin} induces non-trivial correlations between spins in the lattice, and therefore walking defects may interact with each other through the intermediary spins. Exponentially decaying probability distributions of the distance between defect pairs indicate topologically ordered liquid-like phases with exponentially decaying correlations and deconfined topological defects. Even though we adopt the ensemble of maximum entropy, it is possible that the resulting ensemble has more than topological order \cite{Henley11}: \emph{e.g.} constraints \eqref{eq:cPlaq} and \eqref{eq:cSpin} together may so constrain the possibilities in each plaquette that we get a height model, or we could find an emergent long-range ordered phase.

We focus on the square lattice with four different choices of $\mathcal{G}$ and $\mathcal{S}$. The group $\bZ_5$ with $\mathcal{S} = \{\pm1\}$ is a model isomorphic to the well-studied six-vertex model (also known as the ice model \cite{Baxter82}). The reproduction of analytically determined properties of the six-vertex model by our $\bZ_5$ model serves as a benchmark for our simulations, and the quasi-critical behavior of this model is offered as a counterpoint to the corresponding behaviors in topologically ordered models.  Those topologically ordered models are constructed with the following choices of group and spin subset: (i) $\zTwo$ (with $\mathcal{S} = \{a,b,c\}$) is the smallest locally non-trivial topologically ordered model, (ii) $S_3$ (with $\mathcal{S} = S_3 \setminus e$) is the smallest non-Abelian group, and (iii) $\aFive$ (with $\mathcal{S} = \{\text{elements of order 3}\}$) is the smallest non-Abelian simple group, meaning it has no quotient groups (more motivation for $\aFive$ is in \cite{Henley11}).
%

In summary, generalized height models offer us two different ``tunable parameters": the local symmetry of the model, as defined by the full group $\mathcal{G}$; and the subset of allowed spins. Each different choice of spin set $\mathcal{S}$ can be thought of as tuning a family of models \cite{LambertyPapnikolaouHenley}.

\save{Our measurements are based on Monte Carlo simulations of the models with group choices as described in table \ref{tab:SimulatedModels} and periodic lattices of edge length $L$. Define a ``sweep" of single-site updates as $L^2$ consecutive single-vertex updates at randomly selected vertices. Global defect walks are performed one step at a time, with the ratio of single-site sweeps to individual steps in the defect random walk specified by hand (for most of the simulations presented in this paper that ratio was 1 sweep per 100 defect steps). We have performed simulations on lattice sizes ranging from $4^2$ to $256^2$ sites. All measurements quoted in this paper were averaged over sub-trials of length $1\times10^8$ individual measurements. (This count represents defect steps in the case of defect pair distribution measurements, or entire annihilating walks in the case or sector probability measurements.)}

%% file: NA_SecProb.tex

\emph{\underline{Sector Probabilities}}: The defining property of topological order in our models is a degeneracy, in the thermodynamic limit, of the statistical weights (or equivalently the total entrpoy) of the different sectors: hence, we need a means of measuring the relative probabilities of different sectors in an unconstrained ensemble. This is provided by the non-local update moves, since a completed defect walk that traverses a topologically non-trivial loop through our system takes the configuration into a different sector while obeying detailed balance.

The relative probability $P(\Gamma)$ that we would find ourselves in sector $\Gamma = (\gamma_x, \gamma_y)$ after a defect walk then is proportional to the relative number of configurations in a sector. A critical state will have $P(\Gamma)$ converging to different values for different $\Gamma$. In contrast, in topologically ordered models the probability of being in a sector $\Gamma$ at a lattice length $L$ to depend on the class, and for those probabilities to decay to each other as a function of lattice size, \textit{i.e.}
\begin{equation}
    P(\Gamma,L) = P_\infty(\Gamma) + \eta^L P_0(\Gamma) \exp\left( - L / \ell(\Gamma) \right),
    \label{eq:expSP}
\end{equation}
where $\eta=\pm1$ depending on the model.\save{By definition, for a model to be topologically ordered, the number of configurations in each of the $N_s$ sectors must be the same in the thermodynamic limit: $P_\infty(\Gamma) = 1/{N_s}$ for all $\Gamma$.}

\begin{figure}[!h]
    \includegraphics[width = 3.5 in]{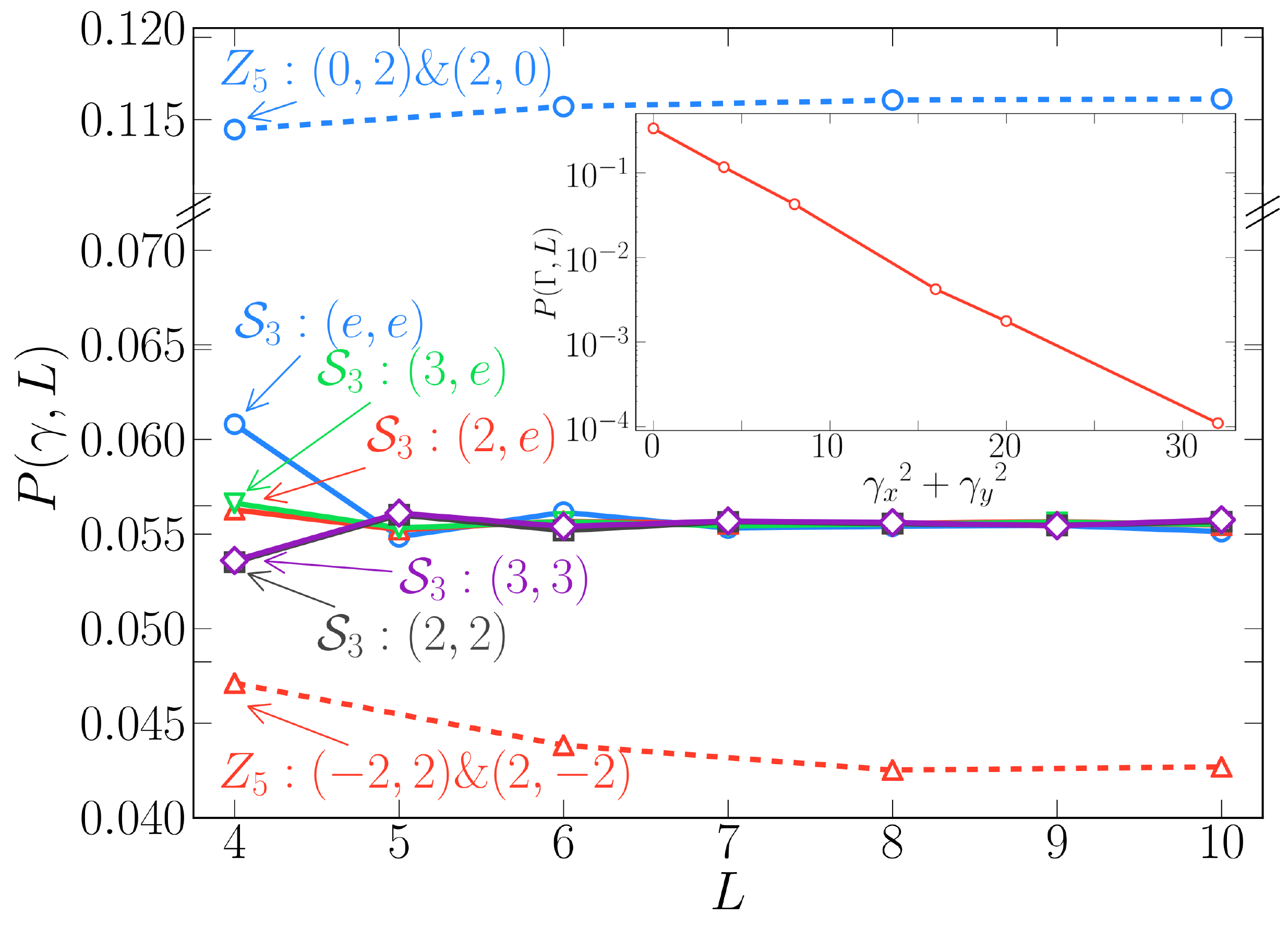}
    \caption{\label{fig:Z5andS3}\textit{Foreground}: Sector probabilities for the model $S_3(2,3)$ (solid lines) and $\bZ_5$ (dashed lines). The sector labels $(\gamma_x,\gamma_y)$ in the inset are integers in $\mathbb{Z}$ labeling the same configurations considered as a 6-vertex model.  For both models, all sectors (originally defined only up to conjugacy) have been further grouped by symmetry ($x$-$y$ symmetry for the ordering of pairs, or any group isomorphism symmetries). \textit{Inset} (log-linear): The exponential scaling of sector probabilities for $\bZ_5$.}
\end{figure}

We start with the benchmark six-vertex model: the number of configurations per sector $\Gamma = (\gamma_x,\gamma_y)$ \cite{footnote1} in the six-vertex model can be calculated analytically \cite{BurtonHenley97}:
\begin{equation}
    P(\Gamma,L) \propto \exp\left(-\dfrac{\kappa}{2}\left({\gamma_x}^2+{\gamma_y}^2\right)\right),\;\kappa=\dfrac{\pi}{6}.
\end{equation}
Note that for integer height models such as this $P(\Gamma,L)$ is \textit{not} dependent on the lattice edge size $L$ \cite{TangSandvikHenley11}. The constancy as a function of $L$ can be seen in figure \ref{fig:Z5andS3}, and the dependence on the sector $\Gamma$ in the inset of that figure. Our simulations produced a value of $\kappa = 0.523(1)$, in agreement with $\kappa = \pi/6$

In figure \ref{fig:Z5andS3} the probability of being in any sector for the model $S_3(2,3)$ converges to a shared constant value at large lattice size $L$, indicating topological order. At short lattice lengths sector probabilities behave differently up to the type of group element (\textit{i.e.} the orders of the elements labeled in the figure; $(2,e)$, $(2,2)$, and $(2,3)$ all have different behaviors). Fit values for many sectors for the topologically ordered models are compiled in table \ref{tab:SectProb}.
\begin{center}
\begin{table}[!h]
    \begin{tabular}{|c|c|c||c|c|c|}
    \hline
    $\Gamma$   &   $P_0(\Gamma)$  &   $\ell(\Gamma)$ &  $\Gamma$   &   $P_0(\Gamma)$  &   $\ell(\Gamma)$ \\
    \hline
    \multicolumn{3}{|c||}{$\mathcal{G} = \zTwo, \mathcal{S}=\mathcal{G} \setminus e$}  &  \multicolumn{3}{c|}{$\mathcal{G} = \aFive, \mathcal{S}=\text{Order 2 elem.}$} \\
    
    \hline
    $(e,e)$     & .3138(8)          & 1.9290(2)         &  $(e,e)$      &  .184(5)          &  1.63(2)          \\
    $(a,e)$     & .0431(1)          & 2.749(4)          &  $(e,2)$      &  .0233(6)         &  1.96(2)          \\
    $(a,a)$     & .01679(9)         & 4.45(2)           &  $(e,3)$      &  .0109(2)         &  2.44(3)          \\
    $(a,a')$    & .1209(4)          & 1.835(3)          &  $(e,5_1)$    &  .0022(8)         &  3.0(5)           \\
    \cline{1-3}
    \multicolumn{3}{|c||}{$\mathcal{G}=S_3, \mathcal{S}=\mathcal{G} \setminus e$}  &  $(e,5_2)$    &  .002(1)  &  3.1(9) \\
    \cline{1-3}
    $(e,e)$     & .45(2)            & 0.838(7)          &  $(2,2)$      &  .048(3)          &  1.06(2)          \\
    $(2,e)$     & .061(7)           & 0.95(3)           &  $(3,3)$      &  .0122(5)         &  1.49(2)          \\
    $(3,e)$     & .053(3)           & 0.96(1)           &  $(5_1,5_1)$  &  .0131(3)         &  2.20(3)          \\
    $(2,2)$     & .097(6)           & 0.98(1)           &  $(5_2,5_2)$  &  .0131(3)         &  2.20(3)          \\
    $(3,3)$     & .20(2)            & 0.86(1)           &  $(5_1,5_2)$  &  .0138(2)         &  2.21(2)          \\
    \hline
    \end{tabular}
    \caption{Fit parameters for the assumed exponential fitting form \eqref{eq:expSP} for Abelian and non-Abelian models. The labels $\Gamma$ are grouped by conjugacy class first and then symmetry (\textit{i.e,} the first and second element represents a conjugacy class, and the whole $(\gamma_x,\gamma_y)$ label represents one or more equivalent conjugacy classes). For example, elements in the labels of sectors for the $S_3$ and $\aFive$ simulations are the order of the conjugacy class ($\aFive$ conjugacy classes $5_1$ and $5_2$ are related by an outer automorphism) and the symmetry grouping is exchange of $x$-$y$ ordering, whereas the elements of labels for the $\zTwo$ model are $a$ for any of the three non-identity elements of that group, and $(a,a')$ when $\gamma_x$ and $\gamma_y$ are not the same element.}
    \label{tab:SectProb}
\end{table}
\end{center}

%% file: NA_DefPairDist.tex

\emph{\underline{Defect Pair Distributions}}: The constraint \eqref{eq:cSpin} generates non-trivial local correlations.  We expect to see exponentially decaying correlations in the topologically ordered models and power law correlations in the $\bZ_5$ model.  For our topologically ordered models, the defects are deconfined at large distances, but have considerable interactions at short ranges.

The distribution function $F_{\delta,\delta'}( \br )$ of the distances between any two defects $\delta, \delta'$ defines an effective entropic potential $V_{\delta,\delta'}( \br )$ via
\begin{equation}
    F_{\delta, \delta'}(\br ) = F_0 \exp\left(-V_{\delta, \delta'}(\br )\right).
    \label{eq:F-V}
\end{equation}

\begin{figure}[ht]
\centering
    \includegraphics[width = 3.5 in]{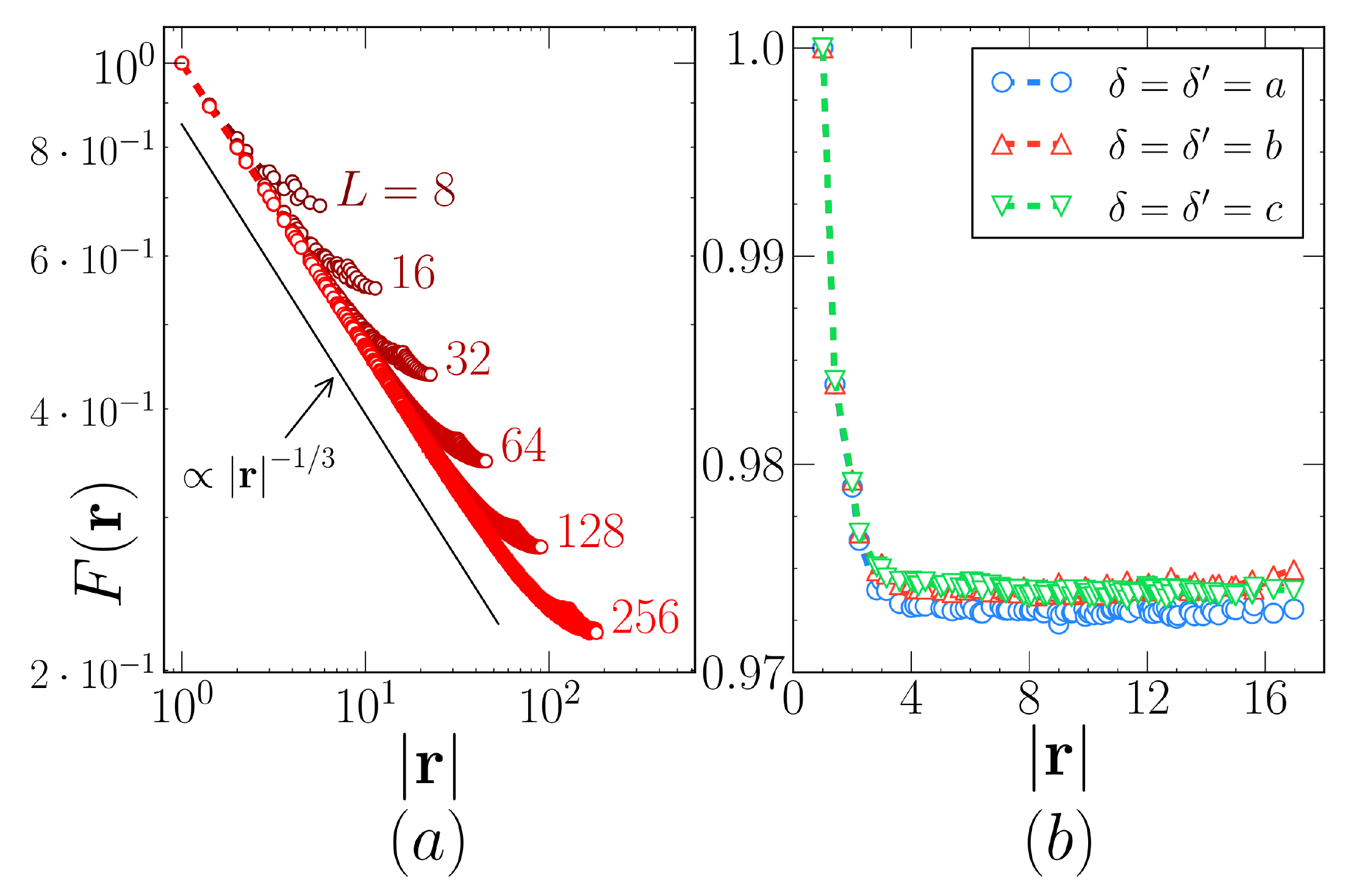}
    \caption{\label{fig:z5andz2} $(a)$ (log-log) the power-law defect pair distribution function $F( \br )$ for the only defect/anti-defect type in the $\bZ_5\{\pm1\}$ model ($+2$ and $-2$) for many lattice sizes $L$. $(b)$ (linear) the defect pair distribution function for the defect types in the $\zTwo$ model (only $L=24$ displayed). Decay to a non-zero constant indicates deconfinement. Note that each of the three distinct \textit{but symmetry-related} defect/anti-defect pairs in the $\zTwo$ model behave equivalently.}
\end{figure}

For the $\bZ_5$ model, we may only have defects of charge $\pm2$, and furthermore it is known that these defects behave like Coulomb charges in two dimensions \cite{Nienhuis87,vanBeijeren77}. The free energy due to their interaction is logarithmic \cite{KondevHenley96}
\begin{equation}
    V_{\delta, \delta'}(\br ) = - \dfrac{\delta \delta '}{2\pi} \kappa \ln( |\br |),
    \label{eq:sixVertexV}
\end{equation}
with $\kappa = \pi / 6$ as noted before \cite{BurtonHenley97}. Hence from \eqref{eq:sixVertexV} the pair distribution follows the power law
\begin{equation}
    F_{\pm 2,\pm 2} (\br ) \propto \exp\left(-\dfrac{1}{3}\ln( |\br |)\right) = {|\br |}^{-1/3},
\end{equation}
where the exponent follows from $\delta = \delta' = \pm2$. Such power-law behavior is indeed seen in the distribution functions for the $\bZ_5$ model shown in figure \ref{fig:z5andz2}$(a)$; the fitted exponent (extrapolated to infinite system size) is $0.330(5)$, in agreement with the expected value of $1/3$.

For topological liquids, exponentially decaying correlations result in a potential which decays exponentially to a constant (defects are \textit{deconfined}). Figure \ref{fig:z5andz2}$(b)$ depicts a rapid decay of defect pair correlations to a non-zero constant, indicating deconfinement.

Analytical transfer matrix toy calculations \cite{Henley11} suggest that the form of \eqref{eq:F-V} ought to be a sum of exponentially decaying terms
\begin{equation}
    F_{\delta,\delta'}(\br ) = F_{\delta,\delta'}^\infty + F_{\delta,\delta'}^0 \exp\left( - |\br | / \xi_{\delta,\delta'} \right).
    \label{eq:dpdExpForm}
\end{equation}
Fit parameters are given in table \ref{tab:dpdExp} for the models which show the exponentially deconfined defect/anti-defect pair distributions characteristic of topological order.

\begin{center}
\begin{table}[!h]
    \begin{tabular}{|c|c|c|c|c|}
        \hline
        $\mathcal{G}$           & $\mathcal{S}$                         & $\delta,\delta'$  & $F_{\delta,\delta'}^0$    & $\xi_{\delta,\delta'}$    \\
        \hline
        $\zTwo$                 & $\zTwo\setminus e$                    & $(a,a)$           & 0.09(2)                   & 0.68(5)                   \\
        \hline
        \multirow{2}{*}{$S_3$}  & \multirow{2}{*}{$S_3 \setminus e$}    & $(2,2)$           & 0.03(4)                   & 0.6(4)                    \\
                                &                                       & $(3,3)$           & 0.01(1)                   & 0.8(4)                    \\
        \hline
    \end{tabular}
    \caption{The parameters for fits of defect pair distribution data to the functional form \eqref{eq:dpdExpForm} indicate exponential decay to a constant background. Fits were also performed for the $\aFive$ model, yielding roughly equivalent values $(F_{\delta,\delta'}^0\approx 0.3$, $\xi_{\delta,\delta'}\approx 0.9)$ for all 10 symmetry-related defect pairs.\save{For the model $\zTwo$ we have averaged all three defect/anti-defect pairs; for $S_3$ we have averaged all pairs of order 2 and order 3 defect/anti-defect pairs, respectively.}}
    \label{tab:dpdExp}
\end{table}
\end{center}

%% file: NA_Conclusions.tex
\emph{\underline{Conclusions}}: Using a generalized definition of topological order, we have explored a new family of classical models which exhibit properties analogous to quantum systems with quantum topological order: (1) topologically robust partitions of massively degenerate ensembles, (2) finite size effects, (3) defect deconfinement (in topological systems), and (4) mediated interactions between charged defects. Elsewhere, we will also exhibit (5) an ability to ``tune" the number of degrees of freedom to transition between topologically ordered and quasi-critical states \cite{LambertyPapnikolaouHenley}. Further studies may illuminate the nature of the interactions between defect/anti-defect pairs, and test a proposed \cite{Henley11} dependence of defect pair interactions on defect charge.

Our generalized height models expand the collection of tractable models that realize topological order, in particular beyond the group $Z_2$ and groups based on it. Various sorts of quantum models could be constructed from these models (either via the Rokhsar-Kivelson construction \cite{RokhsarKivelson88,Henley04,CaselnovoChamonMudryPujol05} or in a manner analogous to the Levin-Wen ``string-net'' construction \cite{LevinWen}).  Whether the non-abelian models contain anyonic defects depends on the phase factors assigned to the matrix elements in quantizing the generalized height model in question.
\save{These models would retain the interesting features of other topologically ordered quantum models (including, possibly, anyonic non-abelian defects), 
but may still be more understandable by virtue of being directly connected to the more tractable classical analogues.}
\save{Do we need to elaborate why $Z_5\{\pm 1\}$ is equivalent to $Z\{\pm 1\}$?}

%% file: NA_Main.bbl
\begin{thebibliography}{99}
\bibitem{XGWen89} X. G. Wen, Phys. Rev. B, \textbf{40}, 7387 (1989).
\bibitem{ThoulessKohmotoNightingale82} D. J. Thouless, M. Kohmoto, M. P. Nightingale, M. den Nijs, Phys. Rev. Lett., \textbf{49}, 405 (1982).
\bibitem{XGWen90} X.-G. Wen and Q. Niu, Phys. Rev. B, \textbf{41}, 9377 (1990).
\bibitem{KalmeyerLoughlin} V. Kalmeyer and R. B. Loughlin, Phys. Rev. Lett., \textbf{59}, 2095 (1987).
\bibitem{WenWilczekZee} X.-G. Wen, F. Wilczek, A. Zee, Phys. Rev. B, \textbf{39}, 11413 (1989).
\bibitem{ReadSachdev} N. Read and S. Sachdev, Phys. Rev. Lett., \textbf{66}, 1773 (1991).
\bibitem{XGWen91} X.-G. Wen, Phys. Rev. B, \textbf{44}, 2664 (1991).
\bibitem{Kitaev03} A. Y. Kitaev, Ann. of Phys., \textbf{303}, 2 (2003).
\bibitem{CastelnovoChamon07} C. Castelnovo and C. Chamon, Phys. Rev. B, \textbf{76}, 174416 (2007).
\bibitem{Henley11} C. L. Henley, J. Phys.: Condens. Matter \textbf{23}, 164212 (2011).
\bibitem{Moessner01} R. Moessner and S. L. Sondhi, Phys. Rev. Lett. \textbf{86}, 1881 (2001).
\bibitem{vanBeijeren77} H. van Beijeren, Phys. Rev. Lett., \textbf{38}, 993 (1977).
\bibitem{BloteHilhorst82} H. W. J. Bl\"{o}te and H. J. Hilhorst, J. Phys. A: Math. Gen., \textbf{15} L631 (1982).
\bibitem{ZhengSachdev94} W. Zheng and S. Sachdev, Phys. Rev. B, \textbf{40}, 2704 (1994).
\bibitem{Levitov90} L. Levitov, Phys. Rev. Lett., \textbf{64}, 92 (1990).
\bibitem{KondevHenley95} J. Kondev and C. L. Henley, Phys Rev B., \textbf{52}, 6628 (1995).
\bibitem{DoucoutIoffe05} B. Dou\c{c}out and L. B. Ioffe, New J. Phys. \textbf{7}, 187 (2005).
\bibitem{Kitaev06} A. Y. Kitaev, Ann. of Phys., \textbf{321}, 2 (2006).
\bibitem{LambertyPapnikolaouHenley} R. Z. Lamberty, S. Papanikolaou, C. L. Henley, unpublished.
\bibitem{Baxter82} Baxter, Rodney J. \textit{Exactly solved models in statistical mechanics.} London: Academic Press Inc. (1982).
\bibitem{footnote1} In the 6-vertex model $(\gamma, \gamma')$ are called ``winding numbers'' and can be any multiple of 2.
\bibitem{BurtonHenley97} J. K. Burton and C. L. Henley, J. Phys. A., \textbf{30}, 8385 (1997).
\bibitem{TangSandvikHenley11} Y. Tang, A. Sandvik, C. L. Henley, Phys. Rev. B, \textbf{84}, 174427 (2011).
\bibitem{Nienhuis87} B. Nienhuis, in {\it Phase Transitions and Critical Phenomena}, edited by C. Domb and J.L. Lebowitz (Academic, London, 1987), Vol. 11.
\bibitem{KondevHenley96} J. Kondev and C. L. Henley, Nucl. Phys. B, \textbf{464}, 540 (1996).
\bibitem{RokhsarKivelson88} D. Rokhsar and S. Kivelson, Phys. Rev. Lett., \textbf{61} 2376 (1988).
\bibitem{Henley04} C. L. Henley, J. Phys.: Condens. Matter \textbf{16} S891 (2004).
\bibitem{CaselnovoChamonMudryPujol05} C. Castelnovo, C. Chamon, C. Mudry, P. Pujol, Ann. of Phys., \textbf{318}, 316 (2005).
\bibitem{LevinWen} M. A. Levin and X.-G. Wen, Phys.Rev. B, \textbf{71}, 045110 (2005).

\end{thebibliography}
